\documentclass[aps,floats,superscriptaddress,floatfix]{revtex4}

\usepackage{epsfig}

\def\beginwide{         \end{multicols} \vspace*{-0.5cm} \noindent
        \rule{3.5in}{.1mm}\rule{.1mm}{5mm} \widetext \medskip }
\def\beginwidetop{
        \end{multicols} \vspace*{-0.5cm} \noindent
        \widetext \medskip }
\def\endwide{
        \hspace*{3.35in}~\rule[-5mm]{.1mm}{5mm}\rule{3.5in}{.1mm}
        \begin{multicols}{2} \vspace*{-1.0cm} \noindent }
\def\endwidebottom{
        \begin{multicols}{2} \vspace*{-1.0cm} \noindent }

\begin{document}
\title{Deblocking of interacting particle assemblies: from pinning to jamming}

\author{M.-Carmen Miguel}
\affiliation{Departament de F\'{\i}sica Fonamental, Facultat de
 F\'{\i}sica, Universitat de Barcelona, Av. Diagonal 647, E-08028, 
Barcelona, Spain}
\author{Jos\'e S. Andrade Jr.}
\affiliation{Departamento de F\'{\i}sica, Universidade Federal do
Cear\'a, 60451-970 Fortaleza, Cear\'a, Brazil.}
\author{Stefano Zapperi}
\affiliation{INFM unit\`a di Roma 1 and SMC, Dipartimento di Fisica,
Universit\`a "La Sapienza", P.le A. Moro 2, 00185 Roma, Italy}
\date{\today}

\begin{abstract}
A wide variety of interacting particle assemblies driven by an
external force are characterized by a transition between a blocked and
a moving phase.  The origin of this deblocking transition can be
traced back to the presence of either external quenched disorder, or
of internal constraints.  The first case belongs to the realm of the
depinning transition, which, for example, is relevant for flux-lines
in type II superconductors and other elastic systems moving in a
random medium. The second case is usually included within the
so-called jamming scenario observed, for instance, in many glassy
materials as well as in plastically deforming crystals. Here we review
some aspects of the rich phenomenology observed in interacting
particle models. In particular, we discuss front depinning, observed
when particles are injected inside a random medium from the boundary,
elastic and plastic depinning in particle assemblies driven by
external forces, and the rheology of systems close to the jamming
transition. We emphasize similarities and differences in these
phenomena.

\end{abstract}

\maketitle

\section{Introduction}

Various materials ranging from synthetic nanocrystals, magnetic
colloids, charged particles in Coulomb crystals, proteins and
surfactants, or vortices in type II superconductors and in
Bose-Einstein condensates, form ordered self-assembled
structures. This topic has attracted much interest for various
fundamental and technological reasons. In this respect, the response
of these structures to external forces of various kinds (optical,
magnetic, mechanical) is of particular importance
\cite{MUR-00,MIT-01,PER-01}. In many cases one observes the presence
of blocked phases, where the evolution of the system is frozen. This
behavior can have different origins: when it is due to the presence of
quenched disorder it is denoted by pinning, while when it is due to
intrinsic constraints it is referred to as ``jamming''. In both
cases, a sufficiently large force leads to a moving phase, through a
deblocking transition.

All these systems can be modeled by a set of interacting particles
moving under the action of external forces sometimes in a random
pinning field. For instance, superconducting vortices in thin films
are pinned by vacancies and driven by an applied current through the
Lorentz force,  colloids interact by Van-der Waals or dipolar forces
and are driven by the solvent flow. Despite the differences
in  these systems, one can try to identify some common features in
their dynamic response. This goal has been achieved mostly through
the use of numerical simulations, which have been extensively used
in the past in various contexts. Here we review the results obtained
from numerical simulations of interacting particles, in order to
provide a common framework for pinning and jamming phenomena that,
despite their similarities, have been traditionally studied by
different communities.

The transition from a blocked to a moving phase is a central problem
in the theory of non-equilibrium critical phenomena. Beside the large
body of theoretical work devoted to the depinning transition of
elastic manifolds in disordered media and recent theories devoted to
jamming in glasses and colloids, one should also mention the theory of
absorbing-state phase transitions \cite{MAR-99,BJP}. An absorbing
state is a configuration in which the evolution of the system,
typically a stochastic lattice model, is frozen and no longer evolves.
When a suitable control parameter is changed the system can eventually
be found into an active, statistically stationary, phase. The
absorbing state phase transition is a second order non-equilibrium
phase transition, characterized by scaling laws and critical
exponents, as in ordinary equilibrium phase transitions.  The same is
true for the depinning transition, and in fact it is sometimes
possible to map a depinning transition into and absorbing-state phase
transition and vice versa. While the characterization of
pinning-depinning as a critical phenomenon is based on a firm
theoretical ground, the current theoretical understanding of jamming
phenomena is not so advanced.  Similarly, when depinning involves the
generation of topological defects, one refers to a plastic depinning
transition, but the precise meaning of the word transition is not
clear at present.

At this stage of the theoretical understanding, however, it is
possible to draw an extensive common picture of these phenomena, in
which some parts are depicted in full detail, others are less precise
and some are just sketched.  We hope that this work, by its own nature
incomplete, will stimulate others to find further connections between
non-equilibrium transitions from blocked to moving phases and possibly
to formulate a complete theory encompassing all these phenomena.

The paper is organized as follows: in section II we discuss the models
used to describe the physics of interacting particle assemblies. In
section III, we analyze the injection of particles in a random medium
and discuss the relations with front propagation and with continuum
theories.  Section IV is devoted to the depinning of interacting
particles by an external force through a pinning field. We analyze the
problem by increasing gradually the level of complexity, from the
pinning of single particle to collective elastic and plastic
depinning. In section V, we introduce jamming phenomena and discuss in
detail the jamming transition observed in plastically deformed
crystals, modeled by a set of stress driven interacting
dislocations. We conclude briefly in section VI.

\section{Interacting particle models and their physical realization}

Several systems in nature can be modeled by a collection of
interacting particles. Here we summarize the main features of these
models and discuss some concrete examples. For simplicity, we will
restrict ourselves to pairwise interactions between particles.  In
this case, identifying the particle coordinates by $\vec{r}_i$, with
$i=1,...N$, we can write the equations of motion in general as
\begin{equation}
m \frac{d^2\vec{r}_i}{dt^2}+\Gamma \frac{d\vec{r}_i}{dt}= 
\sum_{j} \vec{J}(\vec{r}_j-\vec{r}_i)+\vec{F}_{ext}(r_i,t),
\label{eq:gen}
\end{equation}
where $m$ is the mass of the particles, $\Gamma$ is a damping
coefficient, and $\vec {J}= -\vec{\nabla} V(\vec{r})$ is the
interparticle force derived from an interaction potential $V$. The
last term represents external forces, quenched disorder, or other
noise sources, and will be discussed in detail below. In most cases of
interest, dissipation is so strong that we can safely neglect inertia
putting $m=0$.  Most of the following discussion will be devoted to
this {\it overdamped} limit, but occasionally we will discuss as well
the effect of inertia.

Depending of the particular system under study, the interparticle
potential can have different forms which will affect the dynamics of
the system. The simplest case is that of a short-range repulsive
central force $\vec{J}(\vec{r})=\hat{r} K(|\vec{r}|/\xi_v)$ which can
be characterized by its peak value $K(0)$ and its range $\xi_v$. In a
series of increasing complexity, one can consider non-monotonic
interactions (i.e. the force can be repulsive and attractive in
different ranges), long-range interactions (i.e. $K(r) \sim
r^{-\alpha}$ with $0<\alpha<d+2$, for large $r$), anisotropic forces
(i.e. $\vec{J}=\hat{r}K(\vec{r})$), non-central forces
(i.e. $\vec{J}(\vec{r})\times \hat{r} \neq 0$), or different
combinations of the above.

The external force normally includes a uniform driving force $F$,
which could be time dependent. Typical examples are the AC drive
$F(t)=F_0 \sin (\omega t)$ or the ramp up $F(t)=ct$, but the
possibilities are endless. In addition, one should consider position
dependent forces due to quenched impurities that may be present in the
system.  Here we will mainly discuss a set of $N_p$ pinning points placed
randomly at position $R_p$, giving rise to a random force field of the
type
\begin{equation}
F_p(\vec{r})=\sum_p \vec{f}_p((\vec{r}-\vec{R}_p)/\xi_p),
\end{equation}
where $\xi_p$ is the range of the individual pinning forces.  Normally
the particular shape of the pinning potential does not matter as long
as its range is short.  One can also consider correlated disorder,
such as columnar and planar defects, depending on the particular
situation at hand.  In addition, thermal effects can be included
adding a a random uncorrelated Gaussian term $\eta(\vec{r},t)$ to the
equation.

Once the interactions of the particle systems have been specified, one
should also discuss the boundary and initial conditions of the
model. A common choice is to use periodic boundary conditions, and to
place the particles in their zero temperature equilibrium positions
(i.e. forming a crystal).  Alternatively, the particles can be placed
randomly in the system mimicking a sudden quench of the system from a
disordered high temperature phase. The latter may give rise to an
intrinsic geometrical disorder.  These conditions are appropriate if
one is interested in modeling the dynamics in the bulk of the
material, without worrying about surface effects. On the other hand,
in some cases boundary effects are at the core of the phenomena and
one should then implement different initial and boundary
conditions. This case will be discussed explicitly in the next
section.  Periodic boundary conditions need special care when
interactions are long ranged, since in this case one cannot impose a
cutoff to the extent of the interaction force, as it is often done for
short range-forces.  One should instead consider explicitly the
interaction of the particles in a given finite cell with all the
periodic images of the system.  The infinite sum over the images can
rarely be performed exactly and since the sum is slowly converging a
simple truncation of the series gives a poor approximation and may
induce spurious effects. To overcome this problem one can employ the
Ewald summation method, originally proposed for Coulomb interactions,
after generalizing it for the appropriate interactions involved 
\cite{Rapaport}.

Finally, we would like to discuss here some physical realizations of
the generic model we have discussed above.  One of the most studied
examples is the flux-line lattice in type II superconductors. In
superconducting films, the system can be really treated as a two
dimensional set of interacting particles. Conversely, for thicker
superconductor samples one should study a set of flexible lines.
In this article, we will only consider the case of rigid vortex lines.
The interparticle force between rigid flux-lines in the framework of
the London theory is given by
\begin{equation}
\vec{J}(|\vec{r})= \Phi_0^2/(8\pi^2\lambda^3)K_1(|\vec{r}|/\lambda)\hat{r},
\label{eq:bessel}
\end{equation}
where $\Phi_0$ is the quantized flux carried by the vortices, $K_1$ is
a Bessel function and $\lambda$ is the London penetration length
\cite{BLA-94,BRA-95}.  Notice that this is a short-range (since
$K_1(x) \sim \exp(-x)$ for large $x$) repulsive central force, with a
divergence of the form $x^{-1}$ at short distances which is cut off by
the vortex core. Instead in two dimensions, the interaction is
long-range
\begin{equation}
\vec{J}(|\vec{r}|)= \frac{\Phi_0^2\hat{r}}{8\pi^2\lambda^2r},
\end{equation}
decaying as $1/r$.  In addition to interaction forces, a current
$\vec{j}$ flowing in the superconductor produces a Lorentz force
$\vec{F}=\vec{j}\times \vec{B}/c$ acting on the vortices.

In the case of complex fluids or soft condensed matter
materials~\cite{LAR-99}, which contain large polymer molecules or
colloidal particles in a solvent whose molecules are much smaller, a
generic model in which inertial terms are neglected often provides an
effective approach towards describing such systems. The solvent is
considered as a continuum medium, characterized by its viscosity, in
which energy is dissipated as the suspended particles move through
it. In close correspondence with their characteristic dissipative
motion, the suspended particles exhibit a Brownian dynamics due to the
random collisions with the solvent molecules. This is modeled as a
random Gaussian force $\vec{\eta}$ in the equations of motion of the
form

\begin{equation}
\Gamma_i
(\frac{d\vec{r}_i}{dt}-\vec{v}^s)=\vec{J}(\vec{r}_i)+\vec{\eta}(\vec{r}_i,t),
\end{equation}
where $\vec{v}^s$ is the solvent velocity that can be controlled by an
 externally applied flow field, and $\vec{J}$ is an elastic or
 conservative force on particle $i$ due to deformations of long
 molecules or bubbles, or due to other interactions (such as Van der
 Waals, electrostatic, magnetic, and excluded-volume) among the
 suspended particles. The amplitude of the autocorrelation function
 $\langle |\vec{\eta}_i|^2\rangle=2\Gamma_i k_BT$ is proportional to
 the temperature $T$ of the system. More sophisticated algorithms, in
 which one solves similar equations to the one represented above, have
 been developed to model the rheology of dense spherical
 particles~\cite{BRA-88} accounting for hydrodynamic interactions,
 ellipsoidal or rod-like particles~\cite{YAM-94}, as well as
 emulsions~\cite{OHT-90} and foams~\cite{KHA-86}.

Another example which is worth considering from this general point of
view is a collection of dislocations in a thin crystalline
film. Crystal dislocations are topological defects characterized by a
Burgers vector $\vec{b}$~\cite{HIR-92}.  As in the case of
flux-lines, in a three dimensional crystal dislocations are deformable
lines. Nevertheless, one often treats them in the {\em rigid
approximation}, obtaining an effective two dimensional particle model,
which becomes exact for thin samples.  Dislocations produce long-range
stress and strain fields in the host crystal, and experience the
so-called Peach-Koehler force due to the overall local stress.  This
induces an interaction force between dislocations that depends on
their character (edge or screw, when $\vec{b}$ is perpendicular or
parallel to the corresponding dislocation axis,
respectively~\cite{HIR-92}), but that is generally long-range,
decaying as $1/r$, and anisotropic. For instance, the force between
two edge dislocations at a distance $\vec{r}=(x,y)$, and with Burgers
vectors in the $x$ direction is given by
\begin{equation}\label{eq:1}
J_x(x,y)= \frac{\mu b^2}{2\pi(1-\nu)}\frac{x(x^2-y^2)}{(x^2+y^2)^2},
\end{equation} 
where $\mu$ is the shear modulus, and $\nu$ is the Poisson ratio of the
host crystal.  We have only considered the $x$ component since,
differently from flux lines, dislocations move mainly by gliding along
preferential directions, namely the direction of the Burgers
vector. Thus, while the particle system is two dimensional, the motion
is confined along several particular directions. This fact, together
with the anisotropic character of the interaction, gives rise to
metastable structures that act as geometric constrains for their own
dynamics. In this case, the driving force is often an externally
applied stress $\sigma$ which acts on the dislocations through the
Peach-Koehler force $\vec{F}=(\vec{b}\cdot\sigma)\times \hat{L}$,
where $\hat{L}$ is the direction of the dislocation line local
tangent.

\section{Gradient driven dynamics: front invasion}
	
The theoretical and experimental investigation of the growth dynamics
of rough interfaces has became a subject of great scientific interest
in recent years \cite{KAR-86,BAR-95,HAL-95}. This is clearly
illustrated nowadays by the large variety of studies dealing with
front invasion where roughening processes take place such as flow
through porous media \cite{HE-92,RUB-89,STO-88} or imbibition \cite{DUB-00},
flame propagation \cite{ZHA-92,MAU-97}, 
deposition processes \cite{BAR-95,HAL-95}, 
and flux penetration in superconducting materials 
\cite{SUR-99,SUR-98,ZAP-01,MOR-02}.

From a macroscopic point of view, the development of modeling
techniques for the description of these dynamical systems has been
generally based on the traditional approach to transport phenomena,
where the governing expressions are usually differential equations
representing local balances of the quantity of interest (e.g., mass,
momentum, flux of superconducting vortices, etc.) in a {\it continuum}
framework. However, it is sometimes unavoidable that the process of
front propagation takes place on a particular substrate whose
structural details and/or microscopic irregularities cannot be
properly described within a standard macroscopic formalism. On the
other hand, it often happens that these structural features may
represent key factors for the development of highly efficient
materials. This is the case, for example, in the field of
heterogeneous catalysis, where the morphological characteristics of
the catalyst pore space can have a dramatic influence on the
accessibility of the diffusion front of reagent towards the active
sites in the deeper parts of the porous material \cite{AND-97}. 
In the extreme situation where the porous catalyst has a
microscopically disordered, but macroscopically non-homogeneous
geometry, even a departure from the classical diffusion formalism may
be expected \cite{SAH-95}. For example, the so-called {\it anomalous}
type of transport in self-similar (fractal) structures usually occurs
in the form of a subdiffusive regime \cite{HAV-87,HAV-96}.

The invasion of magnetic flux into a disordered type II superconductor
is another problem that has recently been the object of intense
theoretical and experimental research. As a matter of fact, the
magnetization properties of type II superconductors have been studied
for many years, but the interest in this problem has been renewed with
the discovery of high temperature superconductors \cite{BRA-95,BLA-94}.  The
magnetization process is usually described in terms of the Bean model
\cite{BEA-64} and its generalizations: flux lines enter into the sample
and, due to the presence of quenched disorder, give rise to a steady
flux gradient. While the Bean model provides a consistent picture of
average magnetization properties, such as the hysteresis loop and
thermal relaxation effects \cite{KIM-73}, it does not account for local
fluctuations in time and space. It has been recently observed that
flux line dynamics is intermittent, taking place in avalanches
\cite{FIE-95}, and flux fronts are not smooth 
\cite{SUR-99,SUR-98,JAM-00}. In particular, it has been shown 
that the flux front crosses over from flat to fractal as a function 
of material parameters and applied field \cite{SUR-98}.

A widely used modeling strategy to describe the fluctuations around
the Bean state consists numerical simulations of interacting vortices,
pinned by quenched random impurities
\cite{JEN-88,PLA-91,REI-95,REI-96,PLA-96,OLS-97}. With this approach
it has been possible to reproduce flux profiles \cite{REI-95},
hysteresis \cite{REI-95}, avalanches \cite{PLA-91,PLA-96,OLS-97} and
plastic flow \cite{JEN-88,OLS-97}. One of the aims of these studies
\cite{REI-95} is to establish precise connections between the
microscopic models and the macroscopic behavior, captured for instance
by generalized Bean models. A different approach treats the problem at
mesoscopic scale, describing the evolution of interacting
coarse-grained units \cite{BAR-97,BAS-98}, supposed to represent the
system at an intermediate scale.  While these models give a faithful
representation of several features of the problem, the connection with
the underlying microscopic dynamics still represents a very active
field of research.

\subsection{Invading front from an interacting particle simulation}

In this section, the main objective is to show that the gradient
driven dynamics of the overdamped motion of interacting particles in
disordered substrates can display a collective behavior that is typical
of front invasion processes with roughened interfaces. More precisely,
as we show next, the idea is to provide a basic description for these
system in terms of a particle model and to indicate how the relevant
scaling laws relating the front position and the flux profile to the
pinning strength can be consistently extracted from numerical
simulations. In an infinitely long cylinder, the equation of motion
for an interacting particle performing an overdamped motion in a
random pinning landscape can be written as in Eq.~\ref{eq:gen}
\begin{equation}
\Gamma \vec{v}_i = \sum_j \vec{J}(\vec{r}_i - \vec{r}_j)+
\sum_p \vec{f}_p[(\vec{R}_p-\vec{r}_i)/l]+\eta(\vec{r}_i,t),
\label{eq:vf}
\end{equation}
where $\Gamma$ is the effective viscosity, the first term on the right
hand side represents the particle-particle interaction, and the second
accounts for the interaction between particles and pinning centers.
Here we consider that $\vec{J}$ is a short-range interaction, and
$\vec{f}_p$ is the force due to a pinning center, modeled as a localized
trap at the position $\vec{R}_p$, with $\xi_p$ being the range of the
wells (typically $\xi_p \ll \lambda$), and $p = 1, ..., N_p$. 
For example, the pinning force could
be modeled in terms of the expression, $\vec{G}(\vec{x})=-f_0\vec{x}
(|\vec{x}|-1)^2$, for $|\vec{x}|<1$ and zero otherwise. For
completeness, an uncorrelated thermal noise term $\eta$ with zero mean
and variance $\langle \eta^2\rangle = k_b T/\Gamma$ is also added to
the equation, but we will restrict ourselves to the analysis of the
case $T=0$ (see Ref.~\cite{MON-00} for the implementation of thermal
noise in simulations).

\begin{figure}[t]
\centerline{\epsfig{file=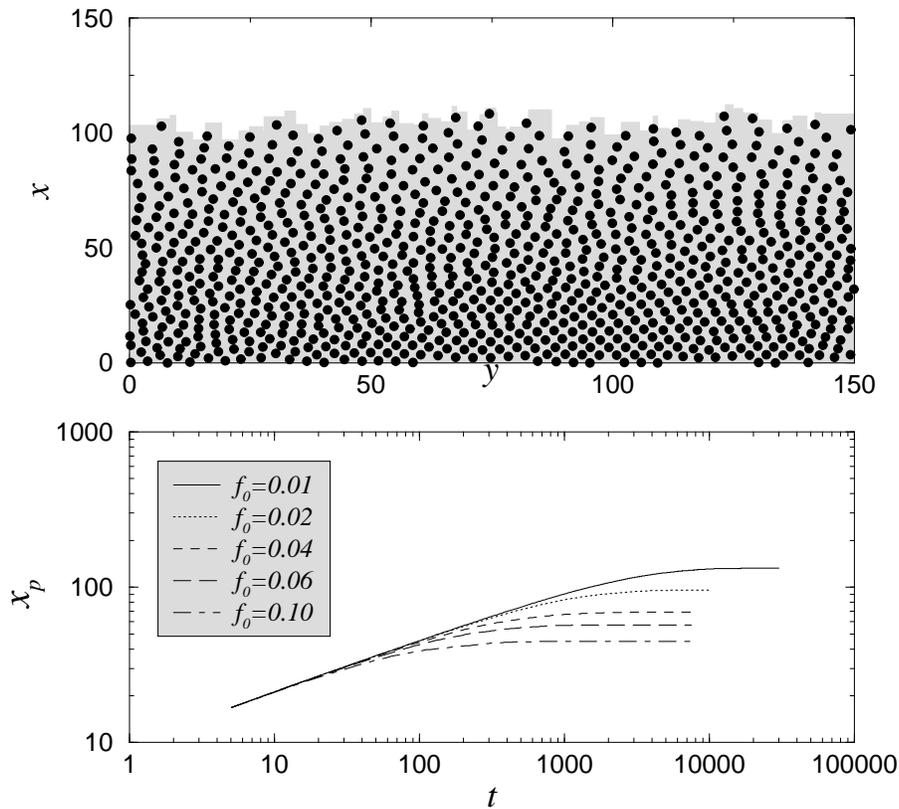,width=12cm,clip=!}}
\caption{In the upper panel we show a typical realization of a flux front
obtained from a simulation of interacting vortices in a disordered landscape
\protect\cite{MOR-02}. In the lower panel, we show the growth of the average 
front position as a function of time for different pinning force strengths.
The front initially grows as $t^{1/3}$ and then is pinned at a distance that  
scales with $f_0$.}
\label{fig:front}
\end{figure}

For gradient driven systems, the solution and interpretation of the MD
model is essentially accomplished by the simultaneous numerical
integration of Eq.~(\ref{eq:vf}) for each moving particle in the
system, and subsequent analysis of the flux front propagation for
different values of the pinning strength $f_0$. For instance, one
typical simulation \cite{MOR-02} can involve up to $N_p=800~000$ Poisson
distributed pinning centers of width $\xi_p=\lambda/2$ in a system of size
$(L_x=800 \lambda , L_y=100\lambda$), corresponding to a density of
$n= 10/\lambda^2$. The number $N$ of flux lines depends essentially on
the boundary condition adopted in the simulation. The injection of
particles into the sample is implemented through the concentration at
$t=0$ of all particles in a small slice $L'\ll\lambda$, parallel to
the $y$ direction, and imposing periodic boundary conditions in both
directions. Due to mutual repulsion in the dense zone of the slice,
the particles will be pushed inside the material, forming a
penetrating front. The front position can then be taken as the $x$
coordinate of the most advanced particle in the system at different times,
or one can divide the system into a grid and identify the front
(see Fig.~\ref{fig:front}).

\subsection{Non-linear diffusion\label{sec:nlin}}
Now we show that the front penetration due to the collective motion of
interacting particles in a substrate of pinning centers can be
described by a disordered non-linear diffusion equation \cite{ZAP-00}.
The equation can be obtained performing a coarse-graining of
Eq.~(\ref{eq:vf}), starting from the Fokker-Planck equation for the
probability distribution of the flux line coordinates
$P(\vec{r}_1,....,\vec{r}_N,t)$
\begin{equation}
\Gamma\frac{\partial P}{\partial t} =
\sum_i \vec{\nabla}_i( -\vec{f}_i P +k_B T\vec{\nabla}_i P),
\label{eq:fp}
\end{equation}
where $\vec{f}_i$ is the force on the particle $i$ given by
Eq.~(\ref{eq:vf}). Next, we introduce the single particle density
$\rho(\vec{r},t)\equiv\langle\sum_i\delta^2(\vec{r}-\vec{r}_i)\rangle$,
where the average is done over the distribution
$P(\vec{r}_1,....,\vec{r}_N,t)$. The evolution of $\rho$ can be
directly obtained from Eq.~(\ref{eq:fp}) and is given by
\begin{equation}
\Gamma\frac{\partial \rho}{\partial t} =
-\vec{\nabla}\left(\int d^2r' \vec{J}(\vec{r}-\vec{r}\;')
\rho^{(2)}(\vec{r},\vec{r}\;',t) 
-\sum_p \vec{f}_p[(\vec{R}_p-\vec{r})/l]\rho(\vec{r},t)\right)
+k_BT \nabla^2\rho,
\end{equation}
where $\rho^{(2)}(\vec{r},\vec{r}\;',t)$ is the two-point density,
whose evolution depend on the three-point density and so on.  The
simplest truncation scheme involves the approximation
$\rho^{(2)}(\vec{r},\vec{r}\;',t)
\simeq \rho(\vec{r},t)\rho(\vec{r}\;',t)$. We then 
coarse grain the equation considering length scales larger
than $\lambda$. This can be done expanding $\vec{J}$
in Fourier space, keeping only the lowest order term
in $\vec{q}$, and retransforming back in real space.
The result reads
\begin{equation}
\int d^2r' \vec{J}(\vec{r}-\vec{r}\;')\rho(\vec{r}\;',t)\simeq 
-a\vec{\nabla} \rho(\vec{r},t),
\label{eq:a}
\end{equation}
where $a\equiv \int d^2r \vec{r}\cdot\vec{J}(\vec{r})/2$.

The coarse graining of the disorder term is more subtle.  A
straightforward elimination of short wavelength modes would give rise,
as in the previous case, to a random force $\vec{F}_c(\vec{r})=
-g\vec{\nabla}n$, where $n$ is the coarse grained version of the
microscopic density of pinning centers $\hat{n}(\vec{r})\equiv
\sum_p\delta^2(\vec{r}-\vec{R}_p)$ and $g \propto f_0$ . This method
can not be applied for short-range attractive pinning forces as the
one we are investigating. In this case, short wavelength modes yield a
macroscopic contribution to pinning that can not be
neglected. Consider for instance the flow between two coarse grained
regions: short-range microscopic pinning forces give rise to a
macroscopic force that should always oppose the motion, while the
random force derived above could in principle point in the direction
of the flow. In other words, $F_c(\vec{r})$ should be considered as a
{\em friction} force whose direction is always opposed
to the driving force $\vec{F}_d$ (in our case $\vec{F}_d=a
\vec{\nabla}\rho$) and whose absolute value is given by
$|g\vec{\nabla}n|$ for $|\vec{F}_d| >|g\vec{\nabla}n|$ and to
$|\vec{F}_d|$ otherwise \cite{CAR-98}.

Collecting all the terms, we finally obtain a disordered non-linear
diffusion equation for the density of particles
\begin{equation}
\Gamma\frac{\partial \rho}{\partial t}=
\vec{\nabla}(a\rho\vec{\nabla}\rho-\rho\vec{F}_c)
+k_BT \nabla^2\rho.
\label{eq:fin}
\end{equation}

Any solution of Eq.~(\ref{eq:fin}) is clearly dependent on the
particular boundary condition (BC) imposed to the system. It is
therefore important to study the effect of different BC's on the
dynamics of front propagation and to show that the macroscopic
approach based on the coarse-grained expression Eq.~(\ref{eq:fin}) is
compatible with the MD model. To be specific, we consider the
situation in which the front penetration takes place in a disordered
type II superconductor and the particles are vortices interacting
according to Eq.~(\ref{eq:bessel}) \cite{MOR-02,ZAP-01}. The following
BC's correspond to different experimental situations \cite{BRY-93,GIL-94}:
\begin{itemize}
\item{(A)} Constant total number of vortices. Experimentally this corresponds 
to an external control of the magnetic flux.
\item{(B)} Constant vortex concentration at the boundary. This case
corresponds to an external control of the magnetic field.
\item{(C)} Total vortex number increasing at constant rate. This 
represents an external control of the flux rate.
\item{(D)} Boundary concentration increasing at constant rate,
corresponding to a constant field rate.
\end{itemize}
One should notice that boundary conditions can be more complicated in
reality, due to complex surface barriers that oppose flux penetration.

For a clean system ($f_0=0$) at $T=0$, Eq.~(\ref{eq:fin}) can be
solved exactly using scaling methods \cite{BRY-93,GIL-94}. In this case,
the density profiles obey the equation
\begin{equation}
\rho(x,y,t)=t^{-\chi} {\cal G}( x/t^{\psi})~,
\end{equation}
where $\chi$ and $\psi$ satisfy $\chi+2\psi=1$. For the BC's
considered: A) $\chi=1/3$, $\psi=1/3$; B) $\chi=0$, $\psi=1/2$; C)
$\chi=-1/3$, $\psi=2/3$; and D) $\chi=-1$, $\psi=1$. These exponents
are in perfect agreement with the results from MD simulations reported
in Ref.~\cite{MOR-02}. For instance, the data reported in
Fig.~\ref{fig:front} correspond to boundary condition (A) and
correctly scale with $\psi=1/3$.  The function ${\cal G}(u)$ also
depends on the boundary condition and for the case (A) is given by
${\cal G}(u)=(1-u^2)/6$ for $u<1$ and vanishes for $u\geq 1$. The
other cases are reported in Refs.~\cite{BRY-93,GIL-94}.

The presence of disorder (pinning centers) induces substantial effects
on the behavior of the system that can be quantified in terms of the
front propagation and/or the shape of the density profiles of flux
lines. Depending on the boundary conditions, it has been observed that
the front is either pinned or simply slowed down \cite{ZAP-01}. Extensive
numerical simulations have also been performed in Refs.~\cite{ZAP-01,MOR-02} 
to show the compatibility between the MD model with disorder and its
coarse-grained representation, Eq.~(\ref{eq:fin}). Moreover, by varying 
the parameters of this continuum description of the front propagation, 
a crossover from flat to fractal flux fronts has been detected, consistent
with experimental observations. The value of the fractal dimension 
suggests that the strong disorder limit is described by percolation. 
In the weak disorder limit, we recover the analytical results derived
in Refs.~\cite{BRY-93,GIL-94}.

\section{Externally driven dynamics: depinning and flow}

In this section, we discuss the response of interacting particles to
an external force in presence of quenched disorder.  The effect of
quenched disorder is first analyzed in a single particle model, which,
although oversimplified, still displays a depinning transition. The
effect of the interparticle potential on the depinning transition will
be first introduced in the framework of the elastic theory, which
breaks down for strong pinning, leading to plastic depinning.
Finally, we briefly discuss the moving phases observed for strong
driving.

\subsection{Single particle pinning}

A first understanding of the dynamics of driven particles
in a disordered landscape can be obtained considering 
the motion of a single particle \cite{GRU-81,FIS-85,ZAP-00}. 
We consider a collection of parabolic wells for the pinning
potentials, so that the equation of motion for a 
single particle is given by
\begin{equation}
\Gamma\frac{d x}{d t}=F + \sum_p f_0 (X_p-x) \theta(|x-X_p|-\xi_p),
\end{equation}
where $x_p$ are the random coordinates of the pinning centers,
which we assume to be non-overlapping.
When the particle is not interacting with a pinning center, it moves
with constant velocity $F/\Gamma$, until it
enters into the attraction range of a pinning center. 
Considering this as the initial condition  (i.e. $x(t=0)= X_p-\xi_p$),
we can solve the equation
\begin{equation}
\Gamma\frac{d x}{dt}=F +f_0 (X_p-x)~~~~ x<X_p+\xi_p,
\end{equation}
which is given by
\begin{equation}
x(t)=x_p+F/f_p-(F/f_0+\xi_p)e^{-t/\Gamma }.
\end{equation}
In the limit $t\to\infty$ the particle remains trapped
as long as  $F< F_c=\xi_p f_0$ and escapes otherwise.
We can interpret this behavior as a depinning transition:
the particle is pinned for $F<F_c$ and moves for $F>F_c$.

The force-velocity diagram can be computed noting that 
for $F> \xi_p f_0$ the particle spends in each trap a time
\begin{equation}
\tau=\Gamma\log((F+F_c)/(F-F_c)).
\end{equation}
The total time $T$ to cross a system of length $L$ with
$N_p$ pinning centers is given by 
\begin{equation}
T = N_p \tau+(L-2\xi_p N_p)\Gamma/F.
\end{equation}
The average velocity of the particle is thus given by
 \begin{equation}
v = \frac{L}{T} = \frac{F}{\Gamma(Fn_p\log((F+F_c)/(F-F
+(1-2\xi_p n_p))},
\end{equation}
where $n_p = N_p/L$ is the number of pinning centers per unit length.
Close to the depinning transition $F_c$ the velocity increases
logarithmically $v \sim 1/\log (F_c-F)$, while at higher forces the
velocity is proportional to the force.  This behavior is
characteristic of the depinning transition, but in general $v\sim
(F-F_c)^\beta$, where $\beta$ is a non trivial critical
exponent. Notice that the logarithmic behavior found above is an
artifact of the discontinuous pinning force. Using a continuous force,
one obtains instead a behavior of the type $v \sim (F-F_c)^{1/2}$
\cite{FIS-85}.  We notice that the dependence of the exponents on the
pinning potential is a peculiarity of the single particle behavior and
is not observed when interactions come into play as we will discuss in
the next section.

Single particle models are also useful for gradient driven dynamics
discussed in the previous section \cite{MOR-02}. In that case the
external force $F$ is replaced by the density gradient, which can be
approximated with a simple function of $x$. For instance, in the case
of boundary condition (B) (see section \ref{sec:nlin}) $F \sim \nabla
\rho \sim \rho_0/x$, where $\rho_0$ is the boundary density. This
simple model yields the correct scaling behavior for the front
position \cite{MOR-02}.
		
\subsection{Elastic depinning}

In the previous section we have only considered the interactions
between moving particles and static pinning centers. We could expect
that those results will be valid only when the interactions between
particles can be neglected, as for instance in a very diluted
system. In most situations, however, interactions between particles
should be explicitly taken into account.  In this case we will still
observe a depinning transition, but its quantitative and qualitative
features will change \cite{KAR-98,NAT-92,NAR-93,LES-97,ERT-94,CHA-00,LED-02}.
It is convenient to first study an interacting particle system in the
{\it elastic} approximation, in which the pinning forces are not
strong enough to break the topological properties of the particle
system.  For instance, if the particles are arranged into a crystal
the external force and the disorder preserve the topological order and
no defects are generated.

In the pinned phase we can write the positions of the particles by
their displacement vectors $\vec{u}(\vec{R}_i)=\vec{r}_i-\vec{R}_i$,
where $\vec{r}_i$ are the coordinates in the deformed system and
$\vec{R}_i$ are the equilibrium positions.  The particle interaction
energy can then be expanded in terms of the displacement field, which
is assumed to be small,
\begin{equation}
U=\sum_{ij}V(\vec{r}_i-\vec{r_j})\simeq U_0+ \sum_{\alpha,\beta}\sum_{ij}
(u_\alpha(\vec{R}_i)-u_\alpha(\vec{R}_j))
\frac{\partial^2 V}{\partial r_\alpha r_\beta}
(u_\beta(\vec{R}_i)-u_\beta(\vec{R}_j)),
\end{equation}
where $U_0$ is the energy in equilibrium. One can then
take the continuum limit expanding in small gradient 
$u(\vec{R})\simeq u(\vec{R}')+
(\vec{R}-\vec{R}')\cdot \vec{\nabla} u(\vec{r})$
and obtain \cite{Ashcroft}
\begin{equation}
U= U_0+ \frac{1}{2}\int d^3r\sum_{\alpha,\beta \gamma \delta} 
E_{\alpha,\beta \gamma \delta} 
\frac{\partial u_\alpha}{\partial r_\gamma}
\frac{\partial u_\beta}{\partial r_\delta},\label{eq:elastice} 
\end{equation}
where the elastic tensor $E_{\alpha,\beta \gamma \delta}$
can be expressed in terms of the interparticle pair potential
as 
\begin{equation}
E_{\alpha,\beta \gamma \delta}=-\frac{1}{2}\sum_{R} R_\gamma R_\delta  
\frac{\partial^2 V}{\partial r_\alpha r_\beta} \label{eq:moduli}
\end{equation}
Clearly this expansion holds as long as the sum in Eq.~\ref{eq:moduli}
converges. If interactions are long range, decaying with a slow power
law $V(r)\sim r^{-(d-1+\sigma)}$ with $\sigma<2$, the elastic energy
can not be expanded in gradients and we have to live with a non-local
interaction.

Coming back to the local limit, a further simplification is obtained
in Eq.~(\ref{eq:elastice}) if we take into account the symmetries of
the equilibrium system.  In the case of an isotropic system, the
elastic tensor has only two independent components and the energy
reduces to
\begin{equation}
U=U_0+ \frac{1}{2}\int d^3r K 
\left(\frac{\partial u_\alpha}{\partial r_\alpha}\right)^2+\mu 
\left(\frac{\partial u_\alpha}{\partial r_\beta}\right)^2,
\end{equation}
where $K$ and $\mu$ are the compression and shear moduli,
respectively.

Using the elastic expression for the interparticle energy, we can
rewrite the equation of motion for the particles as
\begin{equation}
\frac{\partial u_\alpha}{\partial t}=\mu \nabla^2 u_\alpha+
(K+\mu) \frac{\partial}{\partial r_\alpha} (\vec{\nabla}\cdot \vec{u})
+F+f_\alpha(r,u) \label{eq:ela}.
\end{equation}
When expanding around a moving state, we obtain the same equation with
an extra convective term $\vec{v} \cdot \vec{\nabla} u$ on the left
hand side \cite{GIA-96,BAL-97}. These elastic equations are still
impossible to solve exactly, but several results have been obtained
using scaling theories and renormalization group calculations
\cite{KAR-98,NAT-92,NAR-93,LES-97,ERT-94,CHA-00,LED-02} .  In the case
of long-range interactions the gradients are replaced by a non-local
interaction kernel and the equation becomes
\begin{equation}
\frac{\partial u_\alpha}{\partial t}=
\int d^d r' K_{\alpha\beta}(r-r')(u_\beta(r')-u_\beta(r))
+F+f_\alpha(r,u) \label{eq:elalr},
\end{equation}
with $K_{\alpha\beta}(r) \sim 1/r^{d+\sigma}$ for large distances.

A first insight on the behavior of interacting particles in a
disordered media in the elastic approximation can be gained by
collective pinning theory, originally due to Larkin \cite{LAR-79}.  
In this approach, the main energetic contributions entering the problem are
written as a function of a lengthscale $L$.  Minimizing the sum of
elastic and disorder energies, one obtains the characteristic length
$L_c$ at which pinning becomes relevant. Considering a region of size
$L$, where the typical displacements of the elastic medium are of the
order of the range $\xi_p$ of the pinning potential, the elastic energy
can be estimated as $E_{el}\sim\mu L^{d-2}\xi_p^2$, where for simplicity
we have only considered the shear modulus $\mu$. The pinning energy
can be estimated in the limit of weak pinning, when the main effect
comes from the fluctuations in the disorder: in a region of size $L$
the typical fluctuations of the pinning energy scale as $E_{pin}\sim
-E_p \sqrt{L^d n_p}$, where $n_p$ is the density of pins.  
Minimizing the total energy $\mu
L^{d-2}\xi_p^2-E_p \sqrt{L^d n_p}$ with respect to $L$, we obtain the
Larkin length
\begin{equation}
L_c \sim \left(\frac{\mu \xi^2}{\sqrt{n_p} E_p}\right)^{\frac{2}{4-d}}.
\end{equation}
For $L<L_c$ the elastic medium is essentially undeformed, while
elastic deformations grow strongly for $L>L_c$ and typically
the system becomes rough on large lengthscales. 

Notice that the Larkin length decreases with the disorder strength 
as long as $d<d_c=4$.
The dimension $d_c=4$ enters here for the first time, but has in fact
a great importance.  As can be shown by renormalization group $d_c$ is
the upper critical dimension: for $d>d_c$ the system does not roughen
and the critical behavior at the depinning transition is essentially
mean-field like.  Since the physical dimensions is usually lower than
four, one is lead to think that this limit is without practical interest. 
  This is not completely true, since in presence of
long-range interactions $d_c$ typically lowers, reaching sometimes the
physical dimension. In particular, for non local elastic interactions 
with $\sigma <2$ the upper critical dimension is given by $d_c=2\sigma$.

Since $L_c$ represent the characteristic lengthscale at which pinning
start to be effective, Larkin estimated the depinning threshold as the
force needed to unpin a region of length $L_c$.  This can be done
comparing the energy due to the external force $E_{ext}=F\xi_p L^d$ to
the pinning energy at the scale $L_c$, obtaining
\begin{equation}
F_c\sim E_p\sqrt{n_p}L_c^{-d/2}/\xi_p= 
\frac{(f_p\sqrt{n_p})^{4/(4-d)}}{(\xi_p\mu)^{d/(4-d)}}. 
\end{equation}

\begin{figure}[t]
\centerline{\psfig{file=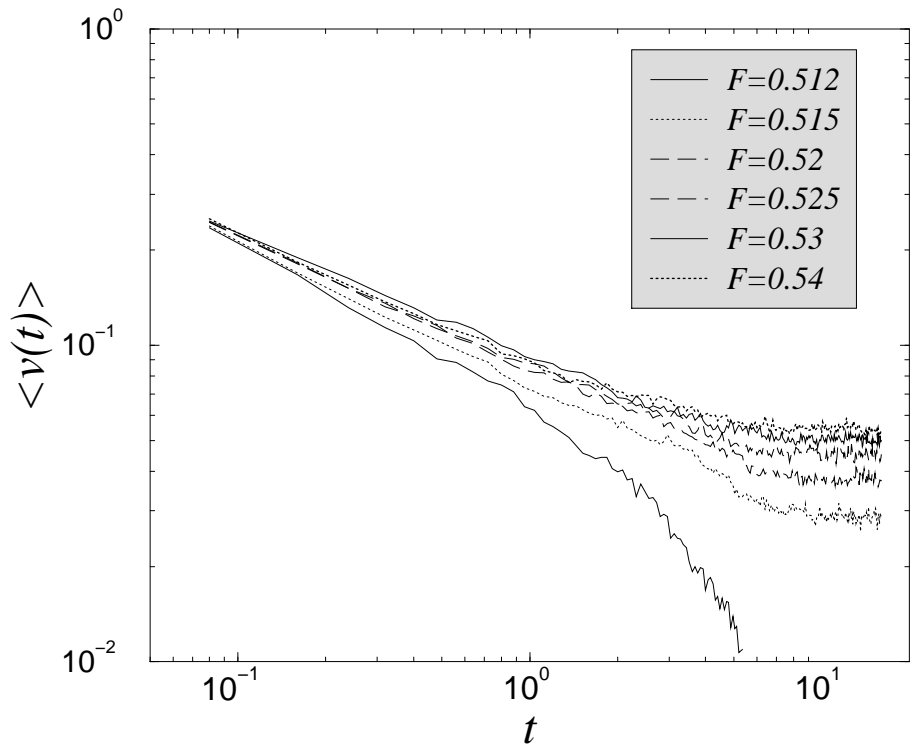,width=12cm,clip=!}}
\caption{The decay of the velocity for $N=400$ interacting particles
with pinning in $d=1$ for different values of the force. 
For $F>F_c=0.514$ the velocity crosses over to a steady value.
See Ref.\protect\cite{ZAP-00} for details on the model.}
\label{fig:vt}
\end{figure}

Collective pinning theory provides a qualitative description of the
depinning transition and a good estimate of the depinning threshold,
but does not allow to estimate the critical exponents and the
force-velocity curve. For this one needs renormalization group
methods, which we do not want to discuss here. We just quote some
results, obtained for the first order in a $\epsilon=4-d$ expansion:
the exponent for the velocity scaling is given by $\beta =1
-\epsilon/9$, while the roughness exponent ruling the scaling of the
typical displacements with the system size $u \sim L^\zeta$ is
estimated as $\zeta=\epsilon/3$.  Other exponents follow from scaling
relations. For instance, one can analyze the transient behavior of the
average particle velocity which initially decays as a power law
$\langle v \rangle \sim t^{-\alpha}$ and then crosses over to a
constant value, when $F>F_c$, or goes exponentially to zero otherwise
(see Fig.~\ref{fig:vt}) \cite{ZAP-00}.  The exponent $\alpha$ can be
obtained by scaling relation as $\alpha=\beta/\nu z$, where $z$ is the
dynamic exponent and $\nu$ is the correlation length exponent
\cite{ZAP-00}.

\begin{figure}[t]
\centerline{\psfig{file=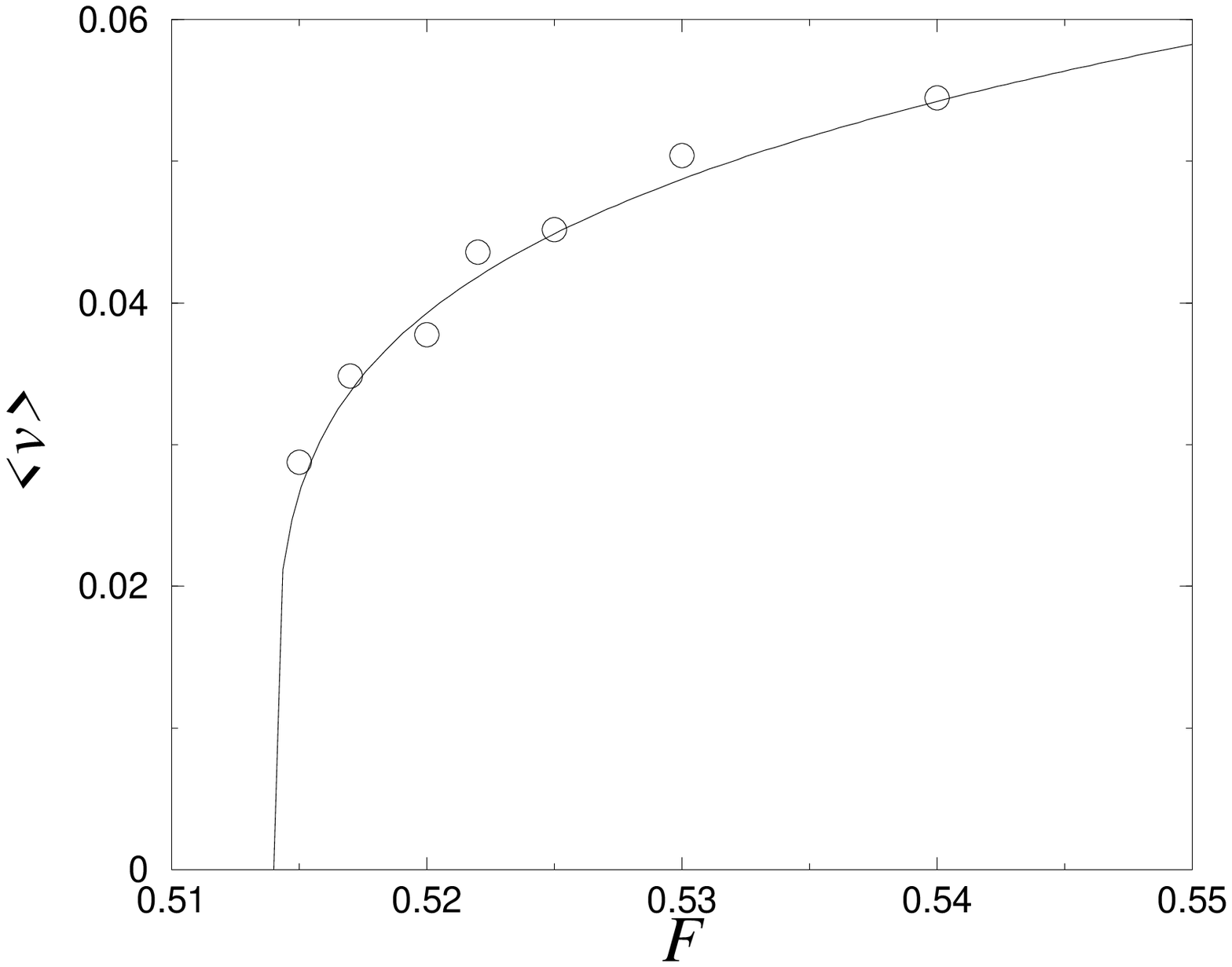,width=12cm,clip=!}}
\caption{The force velocity curve obtained from the steady state velocity
shown in Fig.~\protect\ref{fig:vt}.
with pinning in $d=1$. The best fit yields $\beta=0.22$ and $F_c=0.514$
\protect\cite{ZAP-00}}
\label{fig:elastic_dep}
\end{figure}

Simulations have been widely used in the past to obtain numerically
the value of the exponents. We should distinguish here the vast body
of work pertaining to elastic manifolds, in which the elastic
approximation is enforced directly in the model simulating variants of
Eq.~\ref{eq:ela} \cite{LES-97}, from particle simulations in which the
original system is studied.  In one dimension the elastic
approximation works well and the exponents measured in particle models
reproduce with a good accuracy the results obtained for elastic
manifolds, namely $\beta\simeq 0.25$ and $\zeta\simeq 1.25$ (see
Fig.~\ref{fig:elastic_dep})\cite{ZAP-00,CUL-96,CUL-98,LAC-01}.  This
is not surprising since in $d=1$ topological order is trivially
present. In two dimensions from the mapping with elastic manifold we
would expect $\beta\simeq 0.65$ and $\zeta\simeq 0.75$ \cite{LES-97}
and at least the first exponent is reproduced by particle simulations
\cite{REI-02}.

\subsection{Plastic depinning}

When pinning forces become stronger and/or pinning centers more
diluted the topological order of the system typically breaks down. In
this case it is not possible to describe the deformation in terms of a
displacement field as we did in the previous section and plastic
deformation should be explicitly considered. Due to these
difficulties, a complete theoretical understanding of plastic
depinning is still not available and one should rely on numerical
simulations.  A simple estimate of the conditions for the occurrence
of strong pinning effects can be gained in the framework of collective
pinning theory. When $L_c \sim a_0$, where $a_0$ is the interparticle
distance, the particles are pinned individually.  One can elaborate
these type of arguments and draw phase diagrams in terms of disorder
and applied force \cite{FAL-96}.

\begin{figure}[t]
\centerline{\psfig{file=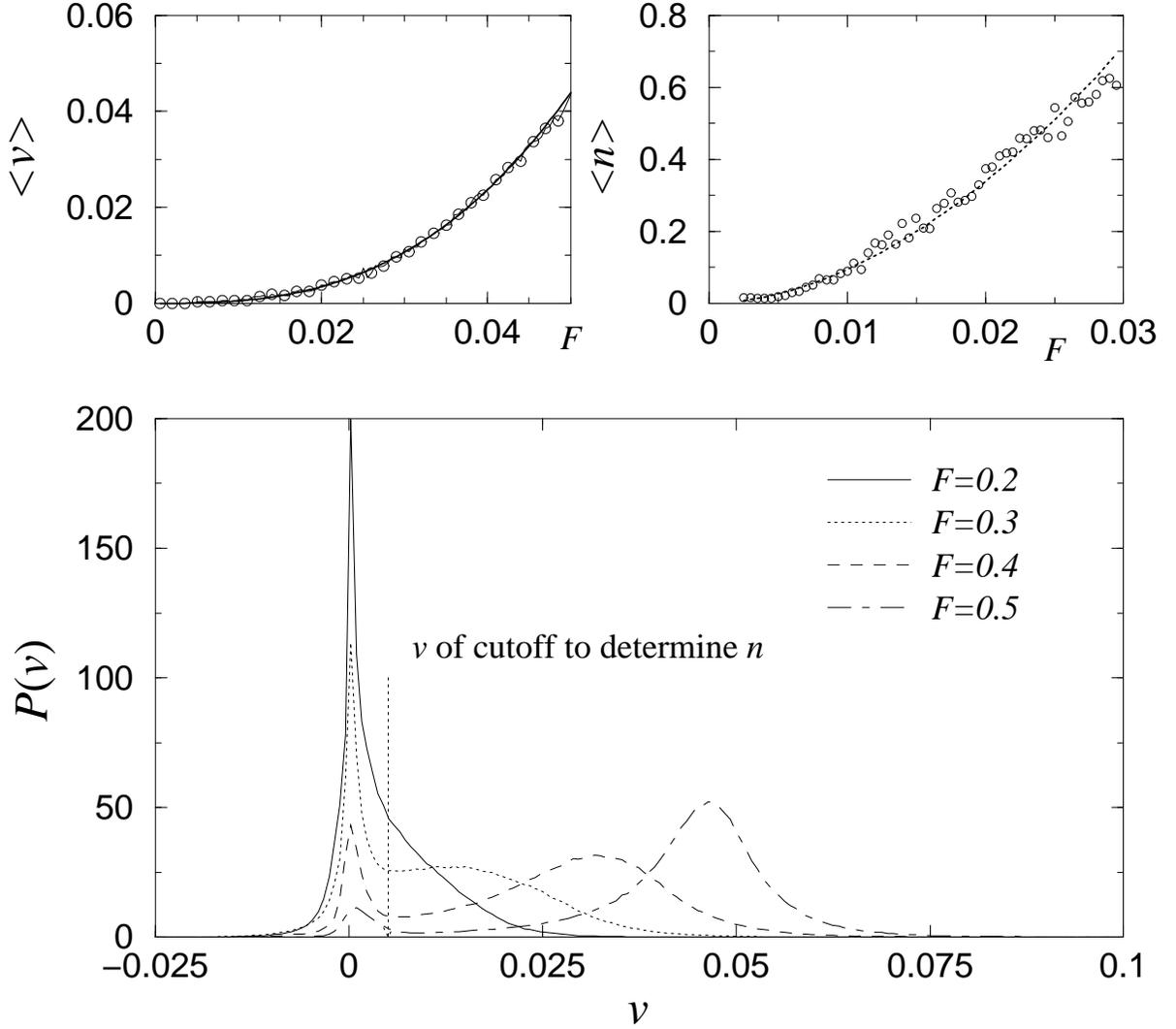,width=16cm,clip=!}}
\caption{(a) The force velocity curve for $N=270$ interacting particles
with pinning in $d=2$ in the strong pinning limit in $d=2$. 
The fit is a power law $F^\beta$ with $\beta=2.7$. 
(b) The velocity distribution for the same system for different values
of the force. Notice the bimodal structure that can be used to set
a threshold and identify moving particles. (c) The average number of
moving particles scaling with the applied force as $F^\gamma$ with
$\gamma=1.8$.}
\label{fig:plastic_dep}
\end{figure}

Here we discuss the main features of plastic depinning as observed in
numerical simulations. Typically the force velocity curve differs
drastically from the one observed when depinning is elastic, which is
upward convex (see Fig.~\ref{fig:elastic_dep}).  In plastic depinning,
the curve is convex downward implying that $\beta >1$. For instance
Ref. \cite{REI-02} finds $\beta=2.2$ in a simulation of
colloids. Others do not provide a value for $\beta$ since it appears
that a precise estimate of the exponent is not straightforward
\cite{FAL-96}. The main reason is that the depinning force $F_c$ is
small and one could as well believe that $F_c=0$ for a large system
and the pinning seen in simulations could be an artifact of finite
sizes.  For instance, the curve reported in Fig.~\ref{fig:plastic_dep}
can be reasonably fit as $v \sim F^\beta$, with $\beta \simeq 2.7$.  A
small value of $F_c$ would not change significantly this fit but it is
difficult to discriminate between the two cases.

The important differences between plastic and elastic depinning can
be highlighted computing the velocity distribution of the moving particles.
In plastic depinning the solid breaks apart: 
some particles are pinned by the strong defects while others move,
typically in channels. Thus the velocity distribution develops two
peaks, one around zero corresponding to pinned particles and
one at higher velocities \cite{FAL-96}. Using this bimodal distribution, it is 
possible to identify the fraction of moving particles $\langle n_m \rangle$,
using a velocity threshold situated in the middle of the two peaks 
(see Fig.~\ref{fig:plastic_dep}).
Again the data are reasonably fit by a power law $F^\gamma$ with 
$\gamma\simeq 1.8$.

\begin{figure}[t]
\centerline{\psfig{file=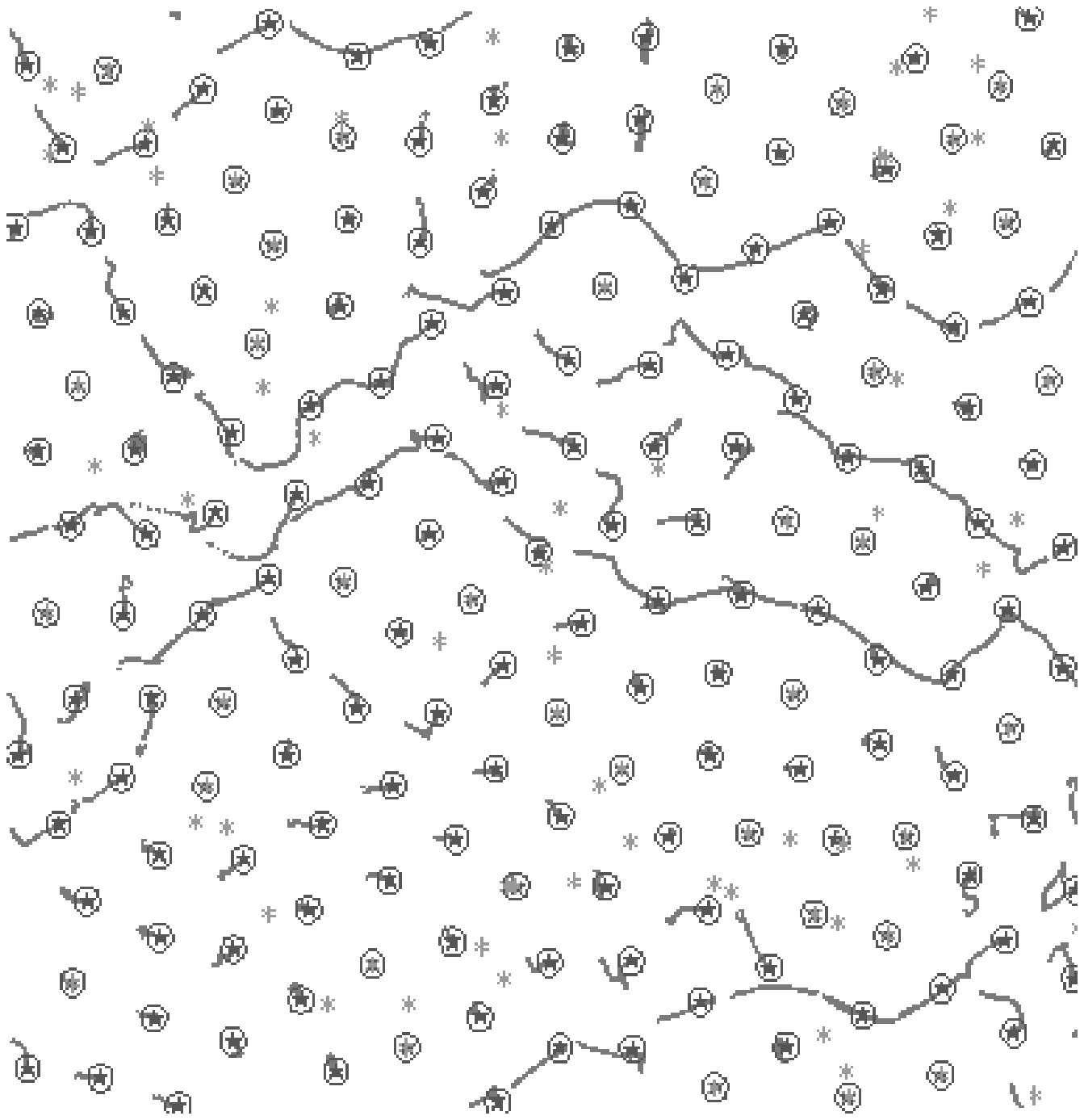,width=8cm,clip=!}}
\caption{An example of particle trajectories in the plastic regime. Particles
are depicted by circles, pinning centers by stars and trajectories by lines.}
\label{fig:pla_tra}
\end{figure}

The channel structure of the dynamics has been studied thoroughly in
the literature analyzing the statistical properties of particle
trajectories \cite{JEN-88,OLS-97,OLS-98}.  In Fig.~\ref{fig:pla_tra}
we report an example of the trajectories in the plastic regime,
showing the separation between pinned and moving particles. It is also
possible to analyze the tearing of the lattice studying its
topological properties (i.e. the presence of defects, such as
dislocations) \cite{FAL-96,OLS-98,FAN-01}.

To understand theoretically plastic depinning several authors have
introduced simplified lattice models, which are possibly amenable to
analytical treatment. For instance, Refs.~\cite{WAT-96,WAT-97} propose
a model in which particles flow in a two dimensional rotated square
lattice.  When the number of particles present in a site is larger
than a threshold, one particle is transferred to a neighboring
site. The flow is directed and the outlet of each site is random but
fixed. It is interesting to note that this model is equivalent to a
sandpile model \cite{BAK-87,DIC-00}, namely a random (quenched)
directed Manna model \cite{MAN-91,PAS-00}.  It should be possible to
use the result obtained for the directed Manna model to solve exactly
this model \cite{PAC-00,KLO-01}.  The model displays a plastic
depinning transition: some particles are trapped and others flow in
channels. The number of particles belonging to the flow basin scales
as $n_m \sim (F-F_c)^\gamma$, with $\gamma \simeq 0.5$. The average
particle current (or velocity) scales instead $v \sim (F-F_c)^\beta$as
$\beta=1.5$. It was conjectured that $\beta=1+\gamma$, implying that
the velocity exponent is due to the combination of the scaling of the
velocity the flowing particle $v_f$ and the one of the channel size.

A similar reasoning was proposed in Ref.~\cite{KAW-99} for plastic
depinning in a disordered XY model. In this model each spin depins as
a single particle in a smooth pinning potential as $v_s \sim
(F-F_c)^{1/2}$. The number of spins depinning also scale with the
applied force as $n_m \sim (F-F_c)^\gamma$, yielding
$\beta=\gamma+1/2$. The exponent $\gamma$ was found to be close to
$d$, the dimension of the lattice, which suggests that a simple
geometrical description could be possible.  We notice that the results
presented in Fig.~\ref{fig:plastic_dep} satisfy approximately the
relation $\beta = 1 + \gamma$, although $F_c \simeq 0$.

Recently, a different approach for plastic depinning was discussed in
Refs.~\cite{MAR-00,MAR-02} through a model of a viscoelastic medium.
In this model depending on the parameter values one observes first
order type or continuous depinning.  The model can be approached by
mean-field theory \cite{MAR-00} and renormalization group
\cite{MAR-02}, but at present the connections with plastic depinning
in particle systems is not clear. It is interesting to notice that a
first order depinning transition, with hysteresis, is also expected in
presence of inertia \cite{PER-98}.  Raising the force from the pinned
phase towards the moving phase is different than decreasing it when
the system is already moving.  In the latter case inertia will keep
the system in motion even beyond the ``depinning threshold'',
resulting in an hysteretic force velocity curve.

To summarize, plastic depinning is characterized in general by
the tearing of the elastic medium through the production of 
dislocations. Only a fraction of the particles move along channels,
while the others are pinned, suggesting to interpret the force velocity curve
by a combination of scaling of single particle velocities and channel
size. However, the wide fluctuations of the numerical values for the exponents
present in the literature and the lack of a theory 
does not allow a clear quantitative picture of the phenomenon.

\subsection{Moving phases, hysteresis and avalanches}

When the external driving force is large enough, particles enter into
a moving phase which can be of several kinds depending on the type of
disorder and the structure of the particle system. All these aspects
are reviewed in Ref.~\cite{GIA-01}.  A first understanding of the
dynamics of an elastic system for strong driving forces comes from a
high velocity expansion \cite{KOS-94}.  When the system flows rapidly
one can write $u(r,t) \simeq Vt + \delta u $, where $V$ is the average
flow velocity. In a first approximation the pinning force becomes
effectively a thermal like noise $f_p(r,u) \simeq f_p(t,Vt)$, with an
effective dynamic temperature decreasing with velocity as $T_d \sim
1/V$.  Thus one would expect that beyond a certain velocity the
elastic system would reorder since the effect of pinning forces
disappear. The discreteness and periodicity of a particle system
modifies considerably this picture: while the displacements along the
direction of motion do not feel the disorder due to the high velocity,
this is not the case for transverse displacements. Taking into account
this effect, Giamarchi and Le Doussal \cite{GIA-96} show that the
system should decouple in elastic channels and predict the existence
of threshold for transverse depinning. These features have been then
observed in simulations \cite{FAN-01}.  For an extensive discussion of
other aspects of the dynamics we refer to Ref.~\cite{GIA-01}

So far we have only discussed the dynamics of interacting particles in
random media occurring under a constant applied force.  When the
applied force is time dependent we observe other interesting
phenomena. In particular, an AC drive leads to hysteresis, which has
been studied in detail for domain walls in ferromagnetic materials
\cite{LYU-99,ZAP-02}, but the results should carry over to a generic
elastic system with disorder \cite{BOC-97}.  Hysteresis is expected in
the quasistatic limit already at the level of a single particle model
driven by an elastic spring \cite{CAR-98}. Close to the depinning
transition, the hysteresis loop of the average displacement as a
function of the force can be obtained analytically solving the
equation \cite{LYU-99}
\begin{equation}
\Gamma\langle du/dt\rangle = (F_0 sin(\omega t)-F_c)^\beta \theta(F-F_c).
\end{equation}
Finally, Ref.~\cite{GLA-03} shows that the velocity force curve also
displays hysteresis, but only in the dynamic regime (i.e. hysteresis
is lost in the limit $\omega \to 0$ when one recovers the usual force
velocity depinning curve).

It is now well established that when the force is raised slowly
towards the depinning threshold the dynamics of an elastic system
takes place in avalanches as in self-organized criticality
\cite{BAK-87}. In particular, the avalanche size distribution scales
as
\begin{equation}
P(s)=s^{-\tau} f(s/s^*), 
\end{equation}
where the cutoff size grows as a power law with the distance from the
critical point $s^* \sim (F_c-F)^{-\nu D}$. Here $\nu$ is the
correlation length exponent and $D$ is the fractal dimension of the
avalanche (i.e. $s^* \sim \xi^D$ ) where $\xi$ is the correlation
length. The connection with self-organized criticality is more
apparent when the elastic system is driven at ``constant velocity''
\cite{LAC-01}.  This can be achieved coupling the system elastically
to a slider moving at constant velocity, or, in other words, replacing
$F$ by $k(Vt - \int d^dr dt \dot{u}(r,t))$.  In the limit of $V\to 0$
and $k \to 0$ the system is driven precisely at the depinning
transition \cite{LAC-01,DIC-00}.  For a general review on avalanches
and self-organized criticality see Ref.~\cite{DIC-00}.

\section{Kinetically constrained dynamics: jamming and shear yielding}

In this section, we will discuss the flow behavior of a wide class of
physical systems whose dynamics is governed by the presence of
kinematical constraints induced by both interactions and geometry. One
of the main features shared by their dynamics is the presence of
jamming, a new concept recently proposed to refer to the suppression
of the temporal relaxation of a physical system and its corresponding
ability to explore the space of
configurations~\cite{LIU-98,LIU-01}. Under the action of externally
applied shear stresses, these systems eventually yield and are able to
flow like a viscous fluid. Shear yielding is thus another feature they
have in common.

As we will see in the following subsections, under stress conditions
both soft materials~\cite{LAR-99} and crystalline solids~\cite{MIG-02}
are susceptible to display jamming and shear yielding due to the
interactions and spatial arrangement of their constituent particles in
the case of soft-matter systems, or to the interactions and spatial
arrangement of their topological defects---such as dislocations---in
plastically deforming crystals. Jamming and yielding could in turn be
responsible for the remarkably similar creep and stationary flow
rheology observed experimentally in these systems, in spite of the big
differences among the materials involved.

\subsection{Jamming and viscoelastic flow in soft condensed matter materials}	

The phenomena of jamming and yielding control the behavior and
properties of soft materials as diverse as colloidal
suspensions~\cite{BON-02}, emulsions~\cite{MAS-96},
foams~\cite{DUR-91}, gels~\cite{SEG-01,BIS-03}, pastes~\cite{CLO-03},
biological tissues~\cite{FAB-01}, and other soft matter systems. Most
of these physical systems consist of various types of soft particles
closely packed into an amorphous state. At such high concentrations,
the individual motion of particles is drastically constrained and, as
a consequence, soft and concentrated materials usually respond like
elastic solids upon the application of low stresses. On the other
hand, they flow like viscous fluids above the so-called yield stress
value $\sigma_y$, exhibiting a common rheology. The deformation
process of amorphous polymeric materials has, for instance, received a
great deal of attention~\cite{LAR-99}. The creep compliance curve of
amorphous polymer networks above the glass transition temperature has
been reported to closely exhibit the following behavior~\cite{FER-80,
PLA-92}
\begin{equation}\label{eq1}
J(t) = \gamma / \sigma = j_0 + C t^{1/3} + t / \eta,
\end{equation}
where $\gamma$ is the global strain of the material, $\sigma$ is the
external stress, $j_0$ is the instantaneous elastic component of the
compliance, and $\eta$ is the viscosity. According to this behavior,
the $C t^{1/3}$ term, also known as Andrade creep term, dominates for
times much smaller than the relaxation time $t_c$ characteristic of
the complex fluid, while a macroscopic viscous flow of the form
$t/\eta$ is established after much longer times $t>>t_c$. One can
define and measure a time dependent effective viscosity in the Andrade
creep regime $\eta=\sigma/\dot{\gamma}\sim t^{2/3}$, which reaches its
equilibrium value at longer times.

A common nonlinear rheology is also observed for higher stress values,
or shear strain-rate values in the case of constant strain-rate
experiments which are in most cases performed in this class of
systems. The stress-strain relationships in the steady state are often
described by phenomenological equations of the form $\sigma=\sigma_y+a
\dot{\gamma}^n$ \cite{BAR-89,FIE-00}, which imply a nonlinear
dependence of the stress $\sigma$ on the shear strain rate
$\dot{\gamma}$. For a Newtonian dispersion $\sigma_y=0$ and $n=1$,
resulting in a constant viscosity coefficient
$\eta=\sigma/\dot{\gamma}$. If instead $\sigma_y\neq 0$, the equation
describes a Bingham fluid. When $n<1$ the relation is known as the
Hershel-Bulkeley law, but if $\sigma_y=0$, the equation describes
a power-law fluid, with a shear strain rate decreasing viscosity.

Although the understanding of yield and viscoplastic flow in these
materials is difficulted by the absence of a clear mediating
mechanism, such as the motion of dislocations in a crystal, it has
been argued that the common rheology displayed by these general class
of complex fluids might be attributed to two particular shared
properties of these materials: structural disorder and metastability;
which are characteristic features of an underlying glassy dynamics
\cite{SOL-97}. In a few words, a glassy dynamics is associated to the
slow structural relaxation which takes place when some parts of the
system are trapped by their neighbors and have to surmount large
energy barriers to explore further more favorable
configurations. Molecular dynamics
simulations~\cite{FAL-98,BAR-01,BER-02,ROT-02} of glass-forming
liquids and polymers have proved of much help in this respect. About
the same time, a general ``jamming'' scenario was also
proposed~\cite{LIU-98} as a common framework to understand the
mechanical behavior of a broader class of non-equilibrium physical
systems (colloidal suspensions, supercooled liquids, foams, etc. and
granular media) which, in spite of their differences, exhibit common
properties such as slow dynamics and scaling features near the
so-called jamming threshold.

Regarding the slow relaxation dynamics characteristic of soft glassy
materials, recent experiments~\cite{KEG-00,WEK-00} show the necessity
of incorporating dynamical heterogeneities for its complete
description. In this respect, a new light scattering method,
introduced in Ref.~\cite{BIS-03}, allowed to detect the intermittent
dynamics of a gel formed from attractive colloids. The dynamics is
found to be intermittent due to random rearrangements which appear to
be localized in time. This and similar experiments strongly suggest
that intermittent behavior seems to be a fundamental ingredient for
the slow relaxation in jammed materials.

Finally, it is interesting to point out that an empirical relation
known as the ``Cox-Merz rule''~\cite{COX-58} quite successfully
relates the non-linear steady rheology with the linear but frequency
dependent rheological properties of polymer melts. In particular, the
Cox-Merz rule relates the steady viscosity at a given shear rate
$\dot{\gamma}$, to the modulus of the dynamic viscosity at a frequency
$\omega=\dot{\gamma}$, i.e. $\eta(\dot{\gamma})=|\eta^*(w)|$. Several
works have {\em a posteriori} tried to theoretically justify this
empirical relation starting from basic
assumptions~\cite{REN-97}. Indeed, the time dependent rheology which
follows from Eq.~(\ref{eq1}) in the Andrade regime
$\eta=\sigma/\dot{\gamma}\sim t^{2/3}$, is dimensionally equivalent to
the shear-thinning behavior $\eta \sim \dot{\gamma}^{-2/3}$ reported
in Ref.~\cite{BER-02} for the steady nonlinear rheology of a binary
Lennard-Jones mixture, and it would be consistent with the Cox-Merz
rule for $w=1/t$.  The shear thinning exponent could also be related
to the subaging behavior observed in many soft-glassy materials $\eta
\sim t_w^{\mu}$, where $t_w$ is the so-called waiting time and $\mu <
1$~\cite{BER-pc}.

In the following section, we will show that most of the attributes
discussed for the case of amorphous soft-glassy materials are also
shared by crystalline materials like soft metals or vortex lattices
deforming plastically due to the motion of dislocations.

\subsection{Dislocation jamming and viscoplastic creep deformation}

\begin{figure}[t]
\centerline{\epsfig{file=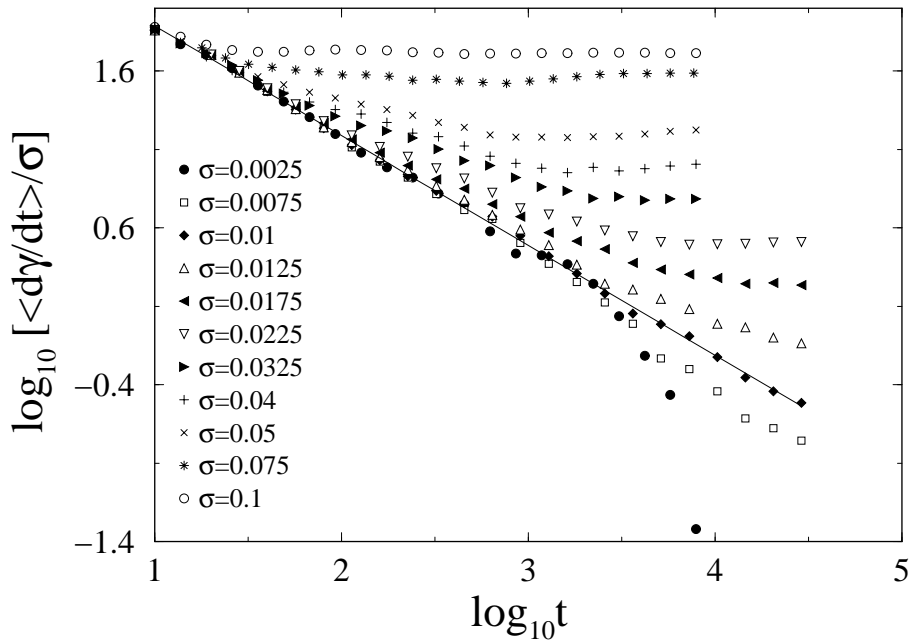,width=12cm,clip=!}}
\caption{The strain rate relaxation for different applied stresses at
$T=0$ for a system of size $L=100b$. The initial density of edge
dislocations is around $1\%$. The solid line is the best linear fit of
the $\sigma=0.01$ curve and yields $d\gamma/dt\sim t^{-0.69}$.}
\label{fig1}
\end{figure}

The viscoplastic deformation of crystalline solids is due to the
creeping motion of dislocations driven by an externally applied
stress~\cite{FRI-67,COT-53,HIR-92,NAB-87}. The study of the dynamics
of these linear topological defects is a subject of considerable
interest because of its practical importance in materials design and
engineering. It is also interesting from the theoretical point of view
for the many features that dislocation motion shares with other
complex systems like, for instance, flux lines in high temperature
superconductors, or some of the soft matter materials discussed above.

At the beginning of the XX century, Andrade reported that the creep
deformation of soft metals at constant temperature and stress grows in
time according to a power law with exponent $1/3$, i.e. $\gamma\sim
t^{1/3}$ where $\gamma$ is the global strain of the material
\cite{AND-10}. More generally, the creep deformation curve usually
follows the relation $\gamma(t)=\gamma_0+ \beta t^{1/3}+\kappa t$,
where $\gamma_0$ is the instantaneous plastic strain, $\beta t^{1/3}$
is known as Andrade creep, and $\kappa t$ is referred to as linear
creep regime~\cite{FRI-67,COT-53}. The same qualitative behavior has
since been observed in many materials with rather different structures
leading to the conclusion that this should be a process determined by
quite general principles, independent of most material specific
properties. Notice that the creep curve for amorphous polymer melts
introduced in the previous subsection, follows exactly the same
relation. Various arguments have been proposed within the dislocation
literature~\cite{FRI-67,COT-53,MOT-53,COT-96,NAB-97} to try to explain
Andrade's creep. Most of them are based on thermally activated
processes over time (or strain) dependent barriers, however, there is
still a lack of consensus on the basic mechanism involved in the
phenomenon.

As in the case of soft-matter systems, the plastic deformation of
crystals only occurs when the externally applied stress overcomes a
threshold value, the yield stress of the material. Above this
threshold value, large-scale dislocation motion may take place, and a
steady regime of plastic deformation is eventually
established. Dislocations tend to move cooperatively under the action
of external stress due to their mutual long-range and anisotropic
elastic interactions, which can be attractive or repulsive. As a
result of these interactions and of the resulting spatial dislocation
structures they give rise to, self-induced constraints build up in the
system and the motion of dislocations may eventually
cease. Nevertheless, small variations of the external loading, the
density, the dislocation distribution or the temperature can enhance
dislocation motion in a discontinuous and intermittent
manner~\cite{MIG-01}.

In Ref.~\cite{MIG-02}, the temporal relaxation of a relatively simple
dislocation dynamics model was studied through numerical
simulation. In particular, a collection of parallel straight edge
dislocations with Burgers vectors $\bf{b}_i$ moving in a single slip
system under the action of constant stress was showed to give rise to
Andrade-like creep at short and intermediate times for a wide range of
applied stresses, without invoking thermally activated processes,
i.e. $T=0$ (see Fig.~\ref{fig1}). The strain rate, which is
proportional to the density of mobile dislocations $d\gamma/dt=\sum_i
b_iv_i$ with $v_i$ the velocity of each dislocation, decays as a power
law with an exponent close to $2/3$, in agreement with Andrade's
observations. At larger times, the strain-rate was observed to cross
over to a linear creep regime (i.e. to a plateau signaling a steady
rate of deformation) whenever the applied stress is larger than a
critical threshold $\sigma_c$, or, otherwise, to decay exponentially
to zero.

These results suggested that a possible interpretation of dislocation
motion and the corresponding creep laws of crystalline materials could
also be found within the general ``jamming'' framework proposed to
encompass the wide variety of non-equilibrium soft and glassy
materials discussed previously. When jammed, these systems are unable
to explore phase space, but they can be unjammed by changing the
stress, the density, or the temperature. The analogies of dislocation
motion and these so-called jammed systems was further explored by
considering the influences of dislocation multiplication, and
thermal-like fluctuations on the dynamics. Dislocation multiplication
favors the rearrangements of the system and induces a linear creep
regime (flowing phase) at lower stress values, but it does not affect
the initial power-law creep. The introduction of a finite effective
temperature $T$ had a similar effect~\cite{MIG-02}.

\begin{figure}[t]
\centerline{\epsfig{file=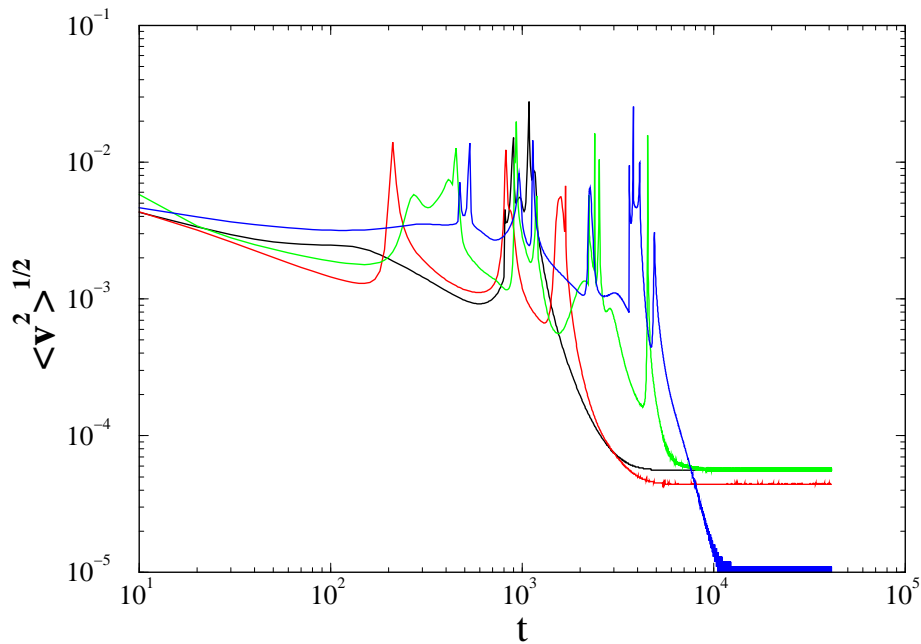,width=12cm,clip=!}}
\caption{The time evolution of root-mean-square velocity of four
single runs of the numerical simulations for different initial
conditions. The applied stress value in all cases is
$\sigma=0.0075$. The curves are depicted in a double logarithmic scale
to emphasize the intermittent bursts characteristic of the creep
dislocation dynamics around the yield threshold.}
\label{fig2}
\end{figure}

The detailed analysis of the model data unveiled the dislocation
microscopic dynamics in the Andrade and in the stationary regimes:
Most dislocations are arranged into metastable structures so that the
stress field they generate in the material is screened out on large
length-scales.  These structures consist of small-angle dislocation
boundaries separating slightly misoriented crystalline blocks or far
more complex dislocation arrangements.  If the applied stress is below
the yield threshold, dislocations are not able to easily explore the
space of configurations to find the most favorable structure
arrangement and they are, most of the time, trapped in metastable
configurations which induce a jamming of the system. Around the yield
threshold, a small fraction of dislocations may, however, attain a
higher mobility and provoke several intermittent rearrangements of the
whole system in the course of time.  The stress field generated by
these unsettled dislocations conserves the initial long-range
character, and forces the system to continue evolving in time in a
cooperative manner to try to reduce the internal shear stress (or
minimize the elastic energy) by exploring further more favorable
arrangements. 

In Fig.~\ref{fig2}, we show the root-mean-square velocity $\langle v^2
\rangle^{1/2}(t)=[\sum_i v_i^2/N]^{1/2}$ of all the dislocations
($N\sim 100-150$) present in a square cell of size $L=100b$ as a
function of time for four single runs of the numerical
simulations. Thus, each run represents the creep behavior of a small
piece (a few nanometers big) of a macroscopic system, and starts from
four different initial dislocation configurations obtained after
letting the system relax in the absence of external load during a
given time interval. The external shear stress applied is in all cases
$\sigma=0.0075$, that is, in the vicinity of the critical threshold
$\sigma_c$. We can clearly appreciate the presence of a few
intermittent burst after which $\langle v^2 \rangle^{1/2}(t)$ slowly
decreases in time. Similar burst, but either positive or negative, can
also be observed in the corresponding strain-rate curves $d\gamma/dt$
(not shown). Andrade's power law creep appears as a result of the
averaging process over many of these runs, mimicking the behavior of a
much bigger system. The closer is the applied stress to the threshold
the longer is the collective power-law motion before the system falls
either in the jammed or in the moving state. Precisely at the critical
point and for the case of an infinite system, the Andrade power-law
could in principle last indefinitely.

\subsection{Non-linear rheology}

\begin{figure}[t]
\centerline{\epsfig{file=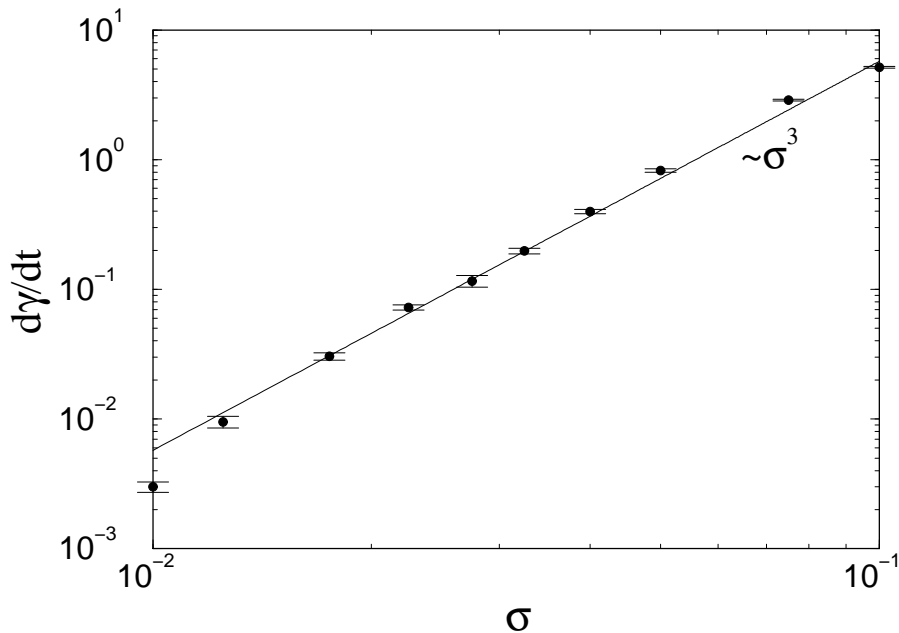,width=12cm,clip=!}}
\caption{The steady strain rate value for different applied stresses
in a double logarithmic scale. The solid line represents a cubic
dependence of the form $d\gamma/dt\sim \sigma^{3}$ which appears to be
in good correspondence with the simulation data for the higher stress
values considered.}
\label{fig3}
\end{figure}

Above the stress threshold, the system eventually exhibits a linear
creep regime in which the dislocations present in the system tend to
glide in a coherent manner. The dependence of the steady strain-rate
value on the external shear stress is shown in Fig.~\ref{fig3}. Within
the error bars, the simulation data for the higher stress values
considered can be fit quite nicely by a cubic law dependence (see the
solid line in the plot). This is an interesting result since, if we
were to compare with the nonlinear rheology characteristic of
amorphous polymeric networks or other soft glassy
materials~\cite{BER-02}, it would correspond to an effective {\em
shear-thinning viscosity} for the dislocation ensemble which decreases
with the strain-rate as $\eta=\sigma (d\gamma/dt)^{-1}\sim
(d\gamma/dt)^{-2/3}$. This result is in good agreement with the
theoretical results obtained in Ref.~\cite{BER-02} and compatible with
the power law shear-thinning behavior $\eta \sim
\dot{\gamma}^{-\alpha}$ with $\alpha=0.5-1.0$ observed in many
different complex fluids~\cite{LAR-99}. 

To summarize, the flow of dislocation structures in crystalline solids
undergoing plastic deformation shares common features with the
time-dependent linear rheology and with the nonlinear steady rheology
of soft glassy materials and, in particular, it seems to satisfy the
empirical Cox-Merz rule. Notice, however, that the concentration of
dislocations in the crystal needs not to be too high to warrant the
presence of kinematical constraints and metastability in the
dynamics. High concentrations could be replaced in this case by the
long-range character of their mutual elastic interactions, that favor
collective motions and rearrangements, and by their ability to form
intricate extended spatial structures (in order to screen out the
stress), that tend to glide in a coherent manner and thus can hamper
their own relative motions driving the system to a jammed
state. Further work is currently under way to try to precisely
identify the most basic mechanism responsible for these remarkable
similarities.

\section{Conclusions}
In this paper we have discussed the collective dynamics of an assembly
of interacting particles and, in particular, the transition from a
blocked to a moving phase.  Transitions of this kind are observed in
different contexts and are due to different mechanisms. When the
particles are blocked by quenched disorder, one typically refers to
the depinning transition, which can be elastic when the medium
preserves its topology through the transition, or plastic when
topological defects, such as dislocations, are generated during the
dynamics.  The driving force for depinning can be due to an externally
applied field, or could be self-generated by a density gradient, as in
the case of front propagation. When the motion is not hindered by
quenched disorder, but by intrinsic constraints one usually refers to
a jamming transition.

Common features of depinning and jamming phenomena are, at the
macroscopic level, the observation of a non-trivial steady-state
force-velocity curve, scaling typically as $v\sim (F-F_c)^\beta$ for
$F>F_c$, and a transient power law relaxation of the velocity $v \sim
t^{-\alpha}$.  At the microscopic level, pinning and jamming systems
are both characterized by a complex energy landscape, with many
metastable states. This leads to an intermittent avalanche-like
response to external perturbations.  Thus despite the different
origins, pinning and jamming have several properties in common which
could be possibly used to construct a comprehensive theory of
deblocking transitions.  Numerical simulations of interacting
particles have played a major role so far to elucidate the detailed
nature of some of these phenomena. The advancement of theoretical
understanding is needed to redirect numerical simulations from a
purely descriptive point of view to a deeper level of analysis.

\section*{Acknowledgements}
We thank A. A. Moreira, J. Mendes-Filho, A. Vespignani and M. Zaiser,
who have contributed to the work reviewed here. We thank H. F. da
Silva for the numerical results reported in
Fig.~\ref{fig:plastic_dep}.  This work is supported by an Italy-Spain
Integrated Action.  MCM is supported by the Ministerio de Ciencia y
Tecnolog\'{\i}a (Spain).  JSA is supported by CNPq.  SZ acknowledges
FUNCAP for supporting his visit to the Universidade Federal do
Cear\'a.

\end{document}